\documentclass[global,twocolumn]{svjour}
\usepackage{latexsym}
\usepackage{graphics}
\journalname{myjournal}
\begin{document}
\title{Waveguide properties of single subwavelength holes demonstrated with radially and azimuthally polarized light}
\author{Jochen M\"{u}ller \and P. Banzer \and S. Quabis\and U. Peschel\and G. Leuchs
\thanks{\emph{E-Mail address:} jmueller@optik.uni-erlangen.de}%
}                     
%
%
\institute{Max-Planck-Research-Group of Optics, Information and Photonics, Erlangen, Germany
\\G\"{u}nther-Scharowsky-Str. 1 / Bau 24, 91058 Erlangen}
\date{Received: date / Revised version: date}
%
\maketitle
\begin{abstract}
We investigate the transmission of focused beams through single subwavelength holes in a silver film. We use radially and azimuthally polarized light, respectively, to excite higher order waveguide modes as well as to match the radial symmetry of the aperture geometry.
Remarkably, the transmission properties can be described by a classical waveguide model even for thicknesses of the silver film as thin as a quarter of a wavelength.
\end{abstract}

\section{Introduction}
\label{intro}
The discovery of enhanced transmission through an array of holes by Ebbesen et al. \cite{Ebbesen} triggered investigations on the transmission of nanoscopic apertures. Understanding respective mechanisms of higher transmission offers the possibility to design holes with prescribed transmission levels. For small apertures this may result in a higher resolution in applications such as near field scanning microscopy and lithography.

In recent years many publications addressed the question of what governs the transmission through small apertures. 
Ebbesen et al. interpreted their measurements as a result of interactions of light and surface plasmon polaritons (SPP), which was confirmed by other groups (\cite{Moreno} and \cite{Ghaemi}). It was shown theoretically that plasmonic effects are not only limited to arrays but can also occur in the case of isolated holes \cite{Chang}.
On the other hand there were theoretical proposals that predicted an enhanced transmission just using waveguide theory \cite{Olkkonen} without any SPP processes being involved. Other authors assumed a hybrid effect of dipole radiation \cite{Popov} or localized waveguide resonances \cite{Ruan} combined with SPP coupling.
A completely different approach attributes the enhancements as well as sup

To clarify how far other mechanisms contribute beyond a wave\-guide theory, it is essential to study the behavior of the corresponding circular modes. When a linearly polarized plane wave is used - as it was the case for all experiments up to now - only the $TE_{11}$ mode contributes to the transmission. Thus it is inevitable to excite higher order modes and study their transmission behavior.  

Our approach is to analyze the next two higher order modes of the circular waveguide, the $TM_{01}$ and $TE_{01}$ mode. To excite them one needs a proper input electromagnetic vector field of radial and azimuthal symmetry, respectively. In addition, the dimensions of the field must match the dimensions of the aperture under study to provide sufficient overlap. This requires strong focusing in the case of a single hole.

\section{Setup and principle of measurement}
\label{sec:setup}
\begin{figure}
\resizebox{0.5\textwidth}{!}{%
  \includegraphics{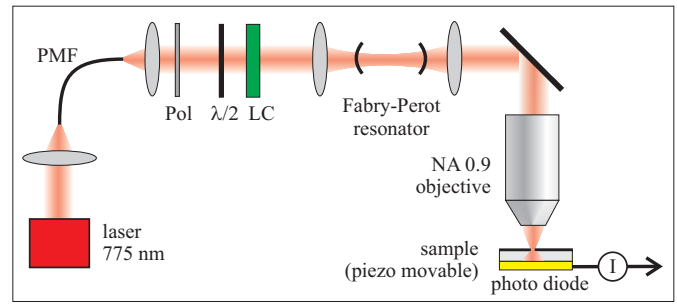}
}
\caption{Scheme of the experimental setup: Light from an ECDL laser is guided through a polarization maintaining fiber (PMF). Using the $\lambda/2$-plate the input polarization for the Liquid Crystal (LC) can be changed switching the output between azimuthal and radial polarization. The Fabry-Perot resonator acts as a mode cleaner. After that the beam is guided by a set of mirrors (only one indicated here) into the microscope objective and is focused onto the sample.}
\label{setup}       
\end{figure}

\begin{figure}
\resizebox{0.5\textwidth}{!}{%
  \includegraphics{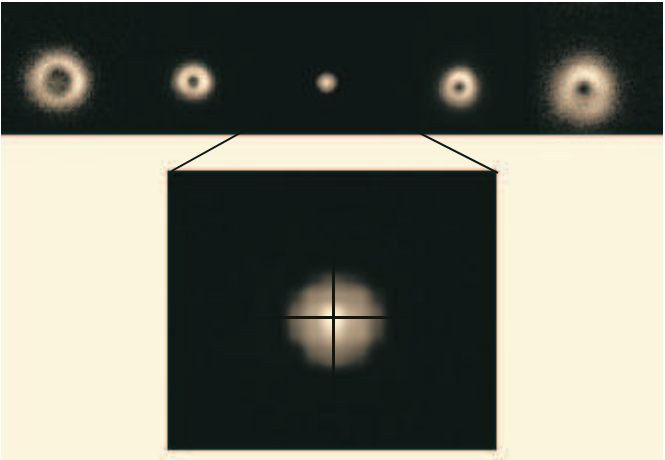}
}
\caption{Principle of measurement: Five transmission images for two positions below the focal plane (left), one in the focal plane (middle) and two above (right) at an increment of 500~nm. Each image was obtained by moving the sample in $x$ and $y$ position and recording the total transmission. The on-axis position (cylindrical symmetric illumination) is then identified by the position which is related to the radial symmetry center of the greyscale pattern (cross-hair picture).}
\label{principle}       
\end{figure}

\begin{figure}
\centering
\resizebox{0.33\textwidth}{!}{%
  \includegraphics{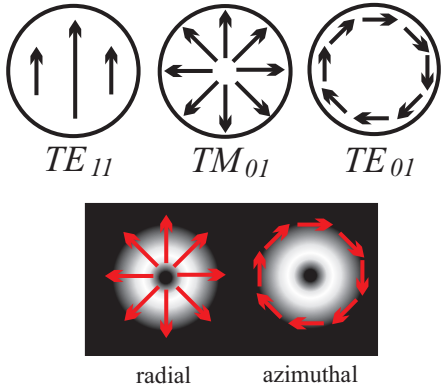}
}
\caption{Transversal field distributions of the relevant modes of an ideal circular waveguide ($TE_{11}$, $TM_{01}$ and $TE_{01}$) compared to those of radial and azimuthal polarization.}
\label{moden}       
\end{figure}
In our experimental setup (see fig. \ref{setup}) the laser beam of an external cavity diode laser (ECDL) operating at 775~nm is cleaned by a polarization maintaining fiber (PMF) for achieving a pure linearly polarized Gaussian mode at the input of the Liquid Crystal cell (LC). The LC converter can produce both azimuthal and radial polarization, depending on the direction of the linear polarization at the input \cite{Stalder}. This direction can be adjusted by a $\lambda/2$-plate. The purity of the radial or azimuthal mode is then enhanced up to 99~\% by a mode-filtering non-confocal Fabry-Perot-resonator which is tuned to be resonant with the desired higher order mode. Finally the beam is focused to a 600 nm FWHM spot by using a high numerical aperture (NA 0.9) objective \cite{Quabis}. 
Silver films were deposited on glass plate substrates (thickness $150~\mu$m) using thermal evaporation. The holes were structured into two films of thickness $t = 110$~nm and $t = 200$~nm using focused ion beam (FIB) technology. The back side of the sample was optically matched to an uncoated bare chip silicon photo diode. With this geometry the diode records the total transmitted light through the apertures within a wide angular spectrum.

Finally, the sample is mounted on a stage which is movable in $x$-$y$-$z$-direction by a piezo system. The motion of the sample is used for adjusting a radially symmetric illumination (on-axis) to the desired structure as well as for precisely identifying the focal plane precisely. 

Plots of the signal of the photo diode as a function of the lateral position of the beam $T(x,y)$ ($z$ close to the focal plane) have radial symmetry (fig. \ref{setup}). The center of the scan pattern refers to on-axis illumination.
The focal position can be identified by taking a series of transmission scans for various values of $z$ around the focal plane. If aberrations are neglected, the scan patterns are symmetric with respect to the focal plane. Furthermore the most compact pattern occurs in the focal plane, since there the size of the light spot is smallest.

\section{Results and discussion}
\label{sec:results}

%
\begin{figure}
\resizebox{0.5\textwidth}{!}{%
  \includegraphics{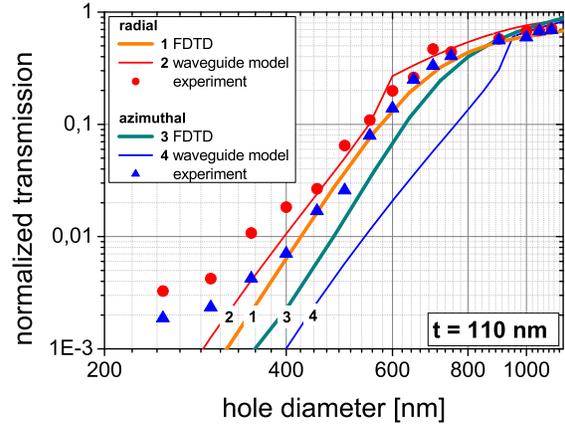}
}
\caption{Experimental data for the transmission through circular apertures etched into a silver film of thickness $t = 110$~nm for radially and azimuthally polarized light (dots) at $\lambda = 775$~nm together with theoretical plots for the ideal waveguide model and rigorous FDTD simulation (solid lines)}
\label{ms110}       
\end{figure}

\begin{figure}
\resizebox{0.5\textwidth}{!}{%
  \includegraphics{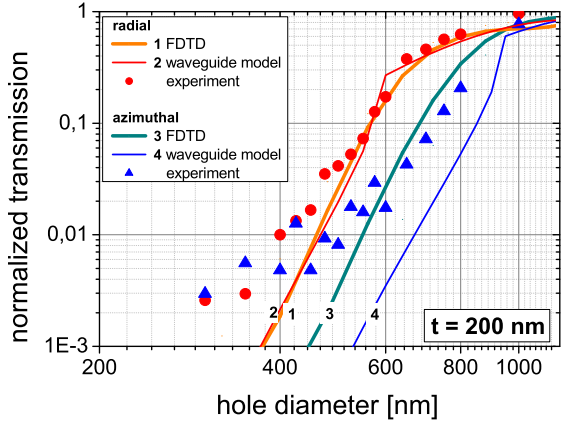}
}
\caption{Same as fig. \ref{ms110}, but with $t = 200$~nm }
\label{ms200}       
\end{figure}

\begin{figure}
\resizebox{0.5\textwidth}{!}{%
  \includegraphics{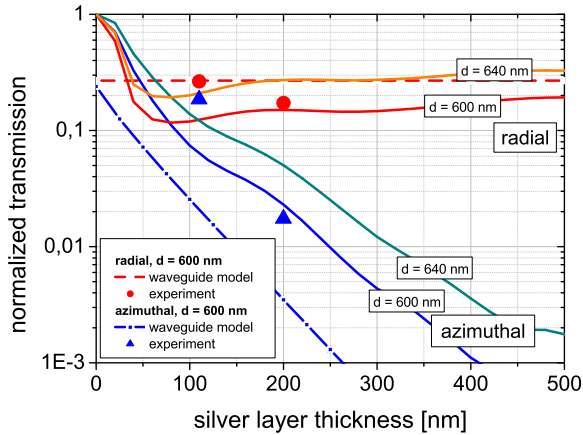}
}
\caption{Experimental data taken for $d = 600$~nm at $\lambda = 775$~nm are displayed. The results of the ideal waveguide model ($d = 600$~nm) and FDTD plots ($d = 600,$ $640$~nm) are plotted as well for varying silver film thickness $t$. The FDTD plots show that the polarization effect tends to disappear at $t = 75$~nm.}
\label{thickvar}       
\end{figure}

Figs. \ref{ms110} and \ref{ms200} present the experimental results for radial and azimuthal polarization. All transmission data have been normalized. The 100~\% transmission reference level was obtained measuring the transmission through a square hole in the silver film with a large edge length of 10~$\mu$m. The noise of the amplified photo current was less than 2~\%. A larger uncertainty arises from not knowing the hole diameters to a precision greater than 30~nm. This results from the clustering properties of the evaporated silver, so the walls of the holes are not homogeneous. Moreover, due to limitations of the etching process, the diameter at the top of the hole differs from that at the bottom.
\\
We applied two theoretical models to the measurements to have a basis for interpreting the data.
The first model assumes the metal to be a perfect conductor. In this case respective waveguide modes are known analytically. The field distribution in the open area of the hole can be decomposed into its circular waveguide eigen modes. Whether or not a mode can travel through the waveguide depends on the mode-specific cut-off diameter

\begin{equation}
d_{c, mn} = \frac{x_{mn}\lambda}{\pi}
\end{equation}

where $x_{mn}$ denotes the zero crossings of the Bessel function $J_{m}$. 

For diameters $d$ larger than the cut-off diameter $d_{c, mn}$ the waveguide is lossless and the excited mode is transmitted without attenuation. If the hole diameter is below cut-off the mode intensity decays exponentially as a function of propagation distance in the waveguide.
As we use radial and azimuthal polarization, the input field at the top of the hole has a high overlap with the $TM_{01}$ and $TE_{01}$ mode, respectively, whereas all other modes including $TE_{11}$ - which is the dominant one for any kind of illumination with linearly polarized light - can be neglected. To our know\-ledge, this is the first time that the transmission through isolated nanoholes is studied experimentally for these two modes. For $d > d_{c, mn}$ both fields are fully transmitted. Nevertheless, the transmission is less than 100 \% because part of the input field energy hits the metal and is reflected.
When the cut-off diameter is reached - 594~nm for the radial $TM_{01}$ mode and 945~nm for the azimuthal $TE_{01}$ mode - attenuation occurs. The larger the thickness the steeper the slope of the transmission curve will be below cut-off.

In fig. \ref{ms200} we can see a significant polarization effect which is shown both by the different transmission values for radial and azimuthal as well as by the waveguide model. Obviously, the waveguide model fits better to the data for the radial polarization than for the azimuthal. This might be a result of the finite conductivity of the silver which allows the field to penetrate. The depth of penetration depends on whether the field lines are perpendicular (in case of radial polarization) or parallel (azimuthal polarization) to the walls of the hole. Therefore the effective hole diameters are different for the two polarization.
As the thickness decreases (see fig. \ref{ms110}) the polarization effect vanishes. Evidently, this cannot be explained by the waveguide model anymore, so a rigorous method such as FDTD is necessary.

For the FDTD simulation we made use of the cylindrical symmetry of the system. With a grid size of 10~nm, the time propagation interval of 0.023~fs was chosen close to the Courant limit \cite{taflove}. We modeled the silver metal using a Drude fit with the parameters $\omega_P = 13.21$~$1/$fs and $\Gamma = 0.1924$~$1/$fs. 
The glass slab as well as the silicon diode were described by lossless dielectric media with the real parts of the indices of refraction 1.5 and 3.6, respectively. 
The FDTD simulation confirms the waveguide model for radial polarization and comes closer to the data for azimuthal polarization than the waveguide model. The reduction of the polarization contrast comparing the results for $t = 110$~nm and $t = 200$~nm is revealed more clearly.
To achieve further insight on how the thickness of the film affects the results we numerically studied a specific hole diameter ($d = 600$ ~nm) and varied the film thickness $t$ (see fig. \ref{thickvar}). For this diameter and our wavelength, the $TM_{01}$ mode propagates without damping. Then, according to the ideal waveguide model, the transmission for the radial polarization is independent of $t$. In contrast, the $TE_{01}$ mode is attenuated because $d = 600$~nm is below its cut-off diameter $d_{c, mn} = 945$~nm, so the transmission for the azimuthal will fade with increasing $t$. Thus the polarization contrast becomes larger with $t$ and vanishes in the limit of $t = 0$. The FDTD simulation, however, shows that the azimuthal transmission is enhanced by about an order of magnitude with respect to the waveguide model. This results in a smaller polarization contrast which already goes to zero at $t = 75$~nm. Qualitatively, this is confirmed by the measurement data. The difference in absolute numbers arises from uncertainties about the actual diameter of the hole. Results obtained with FDTD for another diameter ($d = 640$~nm) close to the reference $d = 600$~nm give an imagination of the sensitivity the transmission with respect to minor changes of the hole geometry.

\section{Summary}
\label{sec:summary}

We found experimentally that radially and azimuthally polarized light is transmitted differently when focusing onto a single subwavelength hole in a silver film of a thickness of about a quarter of the wavelength. It proved to be useful to analyze the problem in the framework of circular waveguide theory while concentrating on two modes with matching polarization symmetry ($TM_{01}$ and $TE_{01}$). The difference in transmission between radially and azimuthally polarized beams becomes small when the thickness is reduced. FDTD simulations show that this reduction with decreasing thickness is more pronounced than predicted by waveguide theory. For thicknesses of around 75~nm and less no significant polarization contrast is observed and expected.

We acknowledge the help of the FIB group of the Fraunhofer-Institut f\"{u}r Integrierte Systeme 
und Bauelementetechnologie (IISB) in Erlangen for structuring the samples.


%

%

\end{document}